\documentclass[aps,prd,onecolumn,groupedaddress,showpacs,nofootinbib,amssymb]{revtex4}
\usepackage[dvips]{graphicx}
\usepackage{amssymb}
\usepackage{amsmath}
\usepackage{graphicx}
\usepackage{amsfonts}
\usepackage{bm}

\begin{document}

\title{Reissner-Nordstr\"{o}m Black Holes in Mimetic $F(R)$ Gravity}

\author{V.K. Oikonomou$^{1,2}$}\,\thanks{v.k.oikonomou1979@gmail.com}
\affiliation{
$^{1)}$ Tomsk State Pedagogical University, 634061 Tomsk \\
$^{2)}$  Laboratory of Theoretical Cosmology, Tomsk State University of Control Systems\\ 
and Radioelectronics (TUSUR), 634050 Tomsk, Russia 
}

\begin{abstract}
In this paper we study under which conditions the Reissner-Nordstr\"{o}m-anti de Sitter black hole can be a solution of the vacuum mimetic $F(R)$ gravity with Lagrange multiplier and mimetic scalar potential. As we demonstrate, the resulting picture in the mimetic $F(R)$ gravity case, is different in comparison to the ordinary $F(R)$ gravity case, with the two descriptions resulting to a different set of constraints that need to hold true. We also investigate the metric perturbations in the mimetic $F(R)$ gravity case, for the Reissner-Nordstr\"{o}m-anti de Sitter black hole metric, at first order of the perturbed variables. Interestingly enough, the resulting equations are identical to the ones corresponding to the ordinary $F(R)$ gravity Reissner-Nordstr\"{o}m-anti de Sitter black hole, at least at first order. We attribute this feature to the particular form of the Reissner-Nordstr\"{o}m-anti de Sitter metric, and we speculate for which cases there could be differences between the mimetic and non-mimetic case. Since the perturbation equations are the same for the two cases, it is possible to have black hole instabilities in the mimetic $F(R)$ gravity case too, which can be interpreted as anti-evaporation of the black hole. 
\end{abstract}


\maketitle

\section{Introduction}

The striking late 90's observations \cite{riess} indicated that the Universe is expanding in an accelerating way. This observation utterly changed our perception for the late-time Universe, with this late-time acceleration being attributed by the scientific community to a negative pressure perfect fluid called dark energy. Modified gravity \cite{fr1,fr2,fr3,fr4,fr5,fr6,fr7,fr8} plays an important role towards the consistent modelling of late-time acceleration, since early-time acceleration \cite{inflation1,inflation2,inflation3,inflation3a,inflation4,inflation5,inflation6,inflation7} can also be described with the same theoretical framework \cite{sergeinojiri,sergeioikonomou1,sergeioikonomou2,sergeioikonomou3,sergeioikonomou4}. For an important stream of reviews and research articles on the concept of dark energy, we refer the reader to \cite{de1,de2,de3,de4,de6,de7}. The latest observational data coming from the Planck telescope collaboration \cite{planck}, indicate that the present time Universe consists of ordinary matter ($\Omega_m\sim4.9\%$), dark energy ($\sim 68.3\%$) and what is perceived as cold dark matter ($\Omega_{DM}\sim 26.8\%$). With regards to the latter, there exist a lot of possible models that can explain dark matter, with most of these assuming that dark matter is described by a particle which does not interact with ordinary matter \cite{shafi,oikonomouvergados}. 

Recently, a quite elegant description of dark matter was given in Ref. \cite{mukh1}, in which the conformal degrees of freedom of the metric in an ordinary Einstein-Hilbert action, can actually mimic dark matter. The approach was given the name mimetic dark matter, and was later further developed in \cite{mukh2,golovnev}. The applications and implications of the mimetic approach are quite many, and has been adopted in many theoretical studies \cite{NO2,mim1,mim3,mim4,mim5,mim6,mim7,mim8,mim9,mim10,mim11,mim12,mim13}. In this paper, we shall be interested in the vacuum $F(R)$ gravity mimetic approach \cite{mim1}, in which case an ordinary vacuum $F(R)$ gravity is equipped with a scalar potential and a Lagrange multiplier \cite{lagra1,lagra2}. Particularly, we shall study in detail for which conditions a Reissner-Nordstr\"{o}m-anti de Sitter (AdS-RN) black hole can be a solution of a general vacuum mimetic $F(R)$ gravity with Lagrange multiplier and scalar mimetic potential. Notice that with the terminology mimetic, we refer to the scalar gravitational degrees of freedom, so we do not have to specify this from now on. The study of Reissner-Nordstr\"{o}m solutions for a general vacuum $F(R)$ gravity was performed in \cite{rnsergei1}, where in order for the Reissner-Nordstr\"{o}m black hole spacetime to be a solution of the corresponding Einstein equations, certain constraints should be satisfied. It is obvious that in the case of mimetic vacuum $F(R)$ gravity, the presence of the mimetic potential and of the Lagrange multiplier will modify the resulting picture, with regards to the constraints that have to be satisfied. Indeed, as we demonstrate, the general set of constraints have differences with the ordinary vacuum $F(R)$ case. For a similar study but for a Schwarzschild-anti de Sitter black hole in vacuum $F(R)$ gravity, see \cite{rnsergei2} and also \cite{antievapsergei7}. Moreover, there exist a large number of studies devoted on black holes solutions in the context of $F(R)$ gravity, and for an incomplete list we refer to \cite{bh1,bh2,bh3,bh4,bh5,bh6,bh7,bh8,bh9,bh10,bh11,bh12,bh13,bh14} and references therein.

The purpose of this paper is twofold: First, we will investigate how the presence of the Lagrange multiplier and of the mimetic potential affects the constraints that need to be satisfied, in order for the AdS-RN metric to be a solution of the vacuum mimetic $F(R)$ gravity. Secondly, we shall study how the aforementioned constraints affect the perturbations of the AdS-RN black hole, studying the problem at first order in the perturbed variables. The resulting picture is quite interesting, since the perturbations of the AdS-RN black hole are not affected by the presence of the Lagrange multiplier and of the mimetic potential. On the other hand, the difference between the ordinary and mimetic $F(R)$ gravity, is that the constraints that must be satisfied in order for the AdS-RN to be a solution of the mimetic $F(R)$ gravity, are different. However, the differences are such so that the resulting perturbation equations are unaffected, at least at first order of the perturbed variables. This feature is probably owing to the specific form of the AdS-RN black hole, but we shall briefly discuss this issue in the end of the paper. Finally, since the perturbation structure of the mimetic AdS-RN black hole, is the same as in the ordinary $F(R)$ gravity case, it is possible to have anti-evaporation \cite{rnsergei1,rnsergei2,hawking,sergeiearly1} of the AdS-RN black hole, in the mimetic case too. 

The outline of the paper is as follows: In section II, we present in brief the mimetic vacuum $F(R)$ gravity formalism with Lagrange multiplier and mimetic potential. We also discuss why the study of the AdS-RN black hole is important and we investigate which constraints have to be satisfied, so that the AdS-RN black hole is a solution of the vacuum mimetic $F(R)$ gravity. As we demonstrate the solutions can be classified in two classes, and we thoroughly investigate only one of these, since the latter leads to the Schwarzschild-anti de Sitter black hole. In section III we investigate how the perturbations of the AdS-RN black hole are affected by the presence of the mimetic potential and of the Lagrange multiplier. Interestingly, the resulting picture is that the perturbations are unaffected by the mimetic potential and Lagrange multiplier, an issue that we briefly discuss in the end of the section. In addition, since the perturbations are unaffected, we demonstrate that, as in the ordinary $F(R)$ case, in the mimetic $F(R)$ case too, the perturbations of the AdS-RN black hole are unstable, a result that can be interpreted as the anti-evaporation of the AdS-RN black hole. Finally, the concluding remarks follow in the end of the paper.

\section{Mimetic $F(R)$ Gravity and Reissner-Nordstr\"{o}m Black Holes}

\subsection{The Mimetic $F(R)$ Gravity Theoretical Framework}

The mimetic $F(R)$ gravity was first studied in \cite{NO2}, and in the context of mimetic $F(R)$ gravity, the conformal symmetry is actually an internal degree of freedom \cite{mukh1}, which is not violated. The mimetic gravity approach was introduced by \cite{mukh1}, and in the context of mimetic gravity, the physical metric $g_{\mu \nu}$ that describes our Universe, can be written in terms of an auxiliary scalar degree of freedom, the scalar field $\phi$,
and also in terms of an auxiliary metric tensor $\hat{g}_{\mu \nu}$, in the following way,
\begin{equation}\label{metrpar}
g_{\mu \nu}=-\hat{g}^{\mu \nu}\partial_{\rho}\phi \partial_{\sigma}\phi
\hat{g}_{\mu \nu}\, .
\end{equation}
The equations of motion are obtained by varying the gravitational action with respect to the auxiliary metric $\hat{g}_{\mu \nu}$ and with respect to the extra scalar degree of freedom, instead of varying the action with respect to the physical metric $g_{\mu \nu}$. From Eq. (\ref{metrpar}), it easily follows that,
\begin{equation}\label{impl1}
g^{\mu \nu}(\hat{g}_{\mu \nu},\phi)\partial_{\mu}\phi\partial_{\nu}\phi=-1\,
.
\end{equation}
As it can be easily verified, the Weyl transformation $\hat{g}_{\mu \nu}=e^{\sigma (x)}g_{\mu \nu}$, leaves Eq. (\ref{metrpar}) invariant, and the auxiliary metric $\hat{g}_{\mu
\nu}$ eventually does not appear in the final action. The Jordan frame mimetic $F(R)$ gravity action, equipped with a scalar field potential $V(\phi)$, and a Lagrange multiplier $\lambda (\phi )$, is equal to \cite{NO2},
\begin{equation}\label{actionmimeticfraction}
S=\int \mathrm{d}x^4\sqrt{-g}\left ( F\left(R(g_{\mu \nu})\right
)-V(\phi)+\lambda \left(g^{\mu \nu}\partial_{\mu}\phi\partial_{\nu}\phi
+1\right)\right )\, .
\end{equation}
In the following, we shall refer to the auxiliary scalar potential $V(\phi)$, as the mimetic potential. In the action of Eq. (\ref{actionmimeticfraction}), we assumed that no matter fluids are present, and thus we study the vacuum mimetic $F(R)$ gravity with mimetic potential and Lagrange multiplier. By varying the action of Eq. (\ref{actionmimeticfraction}), with respect to the physical metric $g_{\mu \nu}$, we obtain the following set of equations,
\begin{align}\label{aeden}
& \frac{1}{2}g_{\mu \nu}F(R)-R_{\mu
\nu}F'(R)+\nabla_{\mu}\nabla_{\nu}F'(R)-g_{\mu \nu}\square F'(R)\\ \notag &
\frac{1}{2}g_{\mu \nu}\left (-V(\phi)+\lambda \left( g^{\rho
\sigma}\partial_{\rho}\phi\partial_{\sigma}\phi+1\right) \right )-\lambda
\partial_{\mu}\phi \partial_{\nu}\phi =0 \, .
\end{align}
Moreover, by varying the action of Eq. (\ref{actionmimeticfraction}),
now with respect to the auxiliary scalar field $\phi$, we obtain,
\begin{equation}\label{scalvar}
-2\nabla^{\mu} (\lambda \partial_{\mu}\phi)-V'(\phi)=0\, .
\end{equation}
Note that the ``prime'' in this case, denotes differentiation with respect to the auxiliary scalar field, but in the rest of the paper, this notation will be used to denote differentiation with respect to the Ricci scalar, unless differently stated. Finally, upon variation of the action (\ref{actionmimeticfraction}) with respect to $\lambda (\phi)$, we get,
\begin{equation}\label{lambdavar}
g^{\rho \sigma}\partial_{\rho}\phi\partial_{\sigma}\phi=-1\, ,
\end{equation}
and by noticing (\ref{lambdavar}), we observe that this result is identical to the one appearing in Eq. (\ref{impl1}). In the following sections we shall extensively use the results of this section, in order to study black holes solutions in vacuum mimetic $F(R)$ gravity.

\subsection{Motivation for Studying the Reissner-Nordstr\"{o}m Black Holes}

Before we start investigating the AdS-RN black holes solutions in the mimetic $F(R)$ gravity, we need to discuss in some detail what is our motivation to study such black hole solutions. The motivation is twofold, since these black hole solutions have applications in the early Universe \cite{sergeiearly1,sergeiearly2,sergeiearly3}, but also have applications in condensed matter systems, via the holographic principle and phase transitions \cite{holo1,holo2,holo3}. 

Particularly, it has been known since the work of Hawking \cite{hawkingevap}, that black holes evaporate and thus effectively their horizon decreases. However, the inverse process for Nariai types black holes is also possible \cite{hawking}. Actually, this anti-evaporation procedure is triggered by instabilities of the perturbations of the Nariai black hole, see Ref. \cite{hawking} for further details on this. The Nariai black holes however, are not black holes that result from the usual gravitational collapse of a star, since these are not asymptotically flat. Therefore, these black holes are relevant for the early Universe only, since these can be primordial black holes of some sort. 

It is remarkable that in the classical $F(R)$ gravity case, anti-evaporation of the Nariai spacetime can occur at the classical level, without any quantum gravity effects being involved \cite{antievapsergei7}. In addition, the Reissner-Nordstr\"{o}m black holes are solutions of the $F(R)$ gravity even in the absence of abelian Maxwell fields \cite{antievapsergei9}. As was demonstrated in \cite{rnsergei1}, the AdS-RN black hole is a solution of the $F(R)$ gravity, if certain constraints are satisfied, and in the present work we shall extend the study of \cite{rnsergei1}, for the case of vacuum mimetic $F(R)$ gravity. Note that the AdS-RN spacetime is similar to the Nariai black hole, so our results can have relevance to the physics of the early Universe, owing to the instabilities of the AdS-RN black hole, which in effect can be responsible for the anti-evaporation of such massive objects.

As we already mentioned, the AdS-RN black hole solutions are relevant for the physics of condensed matter systems, via the holographic principle. Actually, in order to provide a gravitational description of condensed matter phenomena, it is of fundamental importance to find black hole solutions that encompass the physical features of a many-body system and it's corresponding phase diagram. Note that the stability of the system depends of course on the field content of the theory. AdS-RN black holes are relevant in condensed systems study, and these result from a Einstein-Maxwell classical theory as the only static solutions that remain stable below a critical temperature \cite{holo1}. Actually, the ground state of such a system is the extremal AdS-RN black hole, which is the case of black holes studied in this paper. The inclusion of a scalar field in the gravitational action, splits the possible ground states of the system, and the resulting instability of the AdS-RN black hole, actually provides the holographic description of the phase transitions that take place in the dual theory \cite{holo2,holo3}. This kind of phase transition is expected to occur in superfluid or superconducting systems, and their study involves linear perturbations of hairy black holes \cite{holo2,holo3}. Therefore, the presence of an instability in an $F(R)$ gravity AdS-RN black hole, without the presence of a Maxwell field, is rather intriguing to study, since there might be a possible connection to the condensed matter systems, yet to be found.

\subsection{General Study of the Solutions}

In this section, we shall investigate under which conditions, a static metric with constant curvature and spherical symmetry can be a solution of a general vacuum mimetic $F(R)$ gravity. In the following, for convenience, we adopt the notation of Ref. \cite{rnsergei1}. As we already stated, we assume that the spacetime is described by a spherical symmetric and static metric, $g_{\mu \nu}$ with it's line element being of the form, 
\begin{equation}\label{metricstatic}
\mathrm{d}s^2=g_{\mu \nu }\mathrm{d}x^{\mu }\mathrm{d}x^{\nu }=-A(r)\mathrm{d}t^2+B(r)\mathrm{d}r^2+r^2\mathrm{d}\Omega^2\, .
\end{equation}
In Eq. (\ref{metricstatic}), the functions $A(r)$ and $B(r)$ are assumed to be smooth and differentiable functions of $r$, and in addition, $\mathrm{d}\Omega^2$ denotes the metric of unit 2-sphere, that is,
\begin{equation}\label{twosphere}
\mathrm{d}\Omega^2=\mathrm{d}\theta^2+\sin (\theta )^2\mathrm{d}\phi^2\, .
\end{equation}
We shall investigate which conditions much hold true in order for black hole solutions in vacuum mimetic $F(R)$ gravity, to exist, which satisfy the following constraints:
\begin{equation}\label{constraints}
A(r)=\frac{1}{B(r)},\,\,\,R=R_0\, ,
\end{equation}
where $R$ is the Ricci scalar, and $R_0$ a constant. Therefore, we look for solutions which lead to a constant scalar curvature and also for which forms of solutions, the functions $A(r)$ and $B(r)$ satisfy the constraint (\ref{constraints}). The Ricci scalar for the metric (\ref{metricstatic}), is equal to,
\begin{equation}\label{ricciscalar}
R=-A''(r)-\frac{4}{r}A'(r)-\frac{2}{r^2}A(r)+\frac{2}{r^2}\, .
\end{equation}
Notice that for deriving Eq. (\ref{ricciscalar}), we also took into account the constraint (\ref{constraints}), for the function $A(r)$, and the ``prime'' denotes in this case, differentiation with respect to the radial coordinate $r$. Since we assumed that $R=R_0$, we obtain the following differential equation,
\begin{equation}\label{diffeqn}
-A''(r)-\frac{4}{r}A'(r)-\frac{2}{r^2}A(r)+\frac{2}{r^2}=R_0\, ,
\end{equation}
which can easily be solved to yield,
\begin{equation}\label{solvesol}
A(r)=1-\frac{r^2}{12}R_0+\frac{C_1}{r}+\frac{C_2}{r^2}\, .
\end{equation}
By setting $C_1=-M$ and $C_2=Q$, Eq. (\ref{solvesol}) becomes,
\begin{equation}\label{solveeqn1}
A(r)=1-\frac{R_0\,r^2}{12}-\frac{M}{r}+\frac{Q}{r^2}\, .
\end{equation}
Hence, by combining Eqs. (\ref{metricstatic}) and (\ref{solveeqn1}), the metric becomes,
\begin{equation}\label{metricressin}
\mathrm{d}s^2=-\Big{(}1-\frac{R_0\,r^2}{12}-\frac{M}{r}+\frac{Q}{r^2}\Big{)}\mathrm{d}t^2+\frac{1}{\Big{(}1-\frac{R_0\,r^2}{12}-\frac{M}{r}+\frac{Q}{r^2}\Big{)}}\mathrm{d}r^2+r^2\mathrm{d}\Omega^2\, ,
\end{equation}
which is the Reissner-Nordstr\"{o}m-anti-de Sitter black hole spacetime. This black hole solution has two event horizons and one cosmological horizon, only in the case $R_0>2$, which can be easily found by solving the equation $\frac{1}{g_{rr}}=0$. However, for notational convenience, we assume that the event horizons occur at $r=r_0$ and $r=r_1$, so by choosing the parameters $M$ and $Q$ in the way we presented in the Appendix, we have,
\begin{equation}\label{complementary}
A(r)=(1-\frac{r_0}{r})(1-\frac{r_1}{r})\Big{(}1-\frac{(r+r_0)(r+r_1)+r_0^2+r_1^2}{12}R_0\Big{)}\, ,
\end{equation}
from which we can easily see that the two event horizons occur at $r=r_0,r=r_1$ and the cosmological horizon at, 
\begin{equation}\label{stillaliveandfree}
\Big{(}1-\frac{(r+r_0)(r+r_1)+r_0^2+r_1^2}{12}R_0\Big{)}=0\, .
\end{equation}

\subsection{Mimetic $F(R)$ Reissner-Nordstr\"{o}m Black Holes: A Study of the Solutions}

The focus in this section is to investigate under which conditions, the metric of Eq. (\ref{metricressin}) is a solution of the mimetic $F(R)$ gravity equations of motion (\ref{aeden}), with action (\ref{actionmimeticfraction}). We shall search for a vacuum mimetic $F(R)$ solution, meaning that only the mimetic potential $V(\phi )$ and the Lagrange multiplier $\lambda (\phi )$ are present, and no matter fluids are assumed to be present. By combining Eqs. (\ref{scalvar}) and (\ref{lambdavar}), the mimetic $F(R)$ equation of motion of Eq. (\ref{aeden}), can be cast as follows,
\begin{align}\label{equationofmotionbasic}
& \frac{1}{2}g_{\mu \nu}F(R)-R_{\mu
\nu}F'(R)+\nabla_{\mu}\nabla_{\nu}F'(R)-g_{\mu \nu}\square F'(R)+
\frac{1}{2}g_{\mu \nu}\left (-V(\phi)\right )-\lambda
\partial_{\mu}\phi \partial_{\nu}\phi =0 \, ,
\end{align}
which for constant scalar curvature it becomes,
\begin{align}\label{equationofmotionbasicconstantcurv}
& \frac{1}{2}g_{\mu \nu}F(R)-R_{\mu
\nu}F'(R)-
\frac{1}{2}g_{\mu \nu}V(\phi)-\lambda
\partial_{\mu}\phi \partial_{\nu}\phi =0 \, ,
\end{align}
By contracting equation (\ref{equationofmotionbasicconstantcurv}) with the metric $g^{\mu \nu}$, we get the following equation,
\begin{equation}\label{contactedeqn}
2 F(R)-RF'(R)-2V(\phi )-\lambda (\phi )g^{\mu \nu}\partial_{\mu}\phi \partial_{\nu}\phi=0\, ,
\end{equation}
which in view of Eq. (\ref{lambdavar}), leads to the following equation,
\begin{equation}\label{solveqne1}
2F(R)-RF'(R)-2V(\phi )+\lambda (\phi )=0\, .
\end{equation}
By solving Eq. (\ref{solveqne1}), with respect to $F(R)$, we obtain,
\begin{equation}\label{finalfreqn}
F(R)=\frac{RF'(R)}{2}+V(\phi )-\frac{\lambda (\phi)}{2}\, .
\end{equation}
By substituting Eq. (\ref{finalfreqn}) in Eq. (\ref{equationofmotionbasicconstantcurv}), we easily get,
\begin{align}\label{equationofmotionbasicconstantcurvabc}
& \frac{1}{2}g_{\mu \nu}\left ( \frac{RF'(R)}{2}+V(\phi )-\frac{\lambda (\phi)}{2}\right )-R_{\mu
\nu}F'(R)-
\frac{1}{2}g_{\mu \nu}V(\phi)-\lambda
\partial_{\mu}\phi \partial_{\nu}\phi =0 \, ,
\end{align}
which after some simple algebraic manipulations, it becomes,
\begin{equation}\label{simlifiedeqnsmotion}
\left (  \frac{g_{\mu \nu}R}{4}- R_{\mu
\nu}  \right )F'(R)-g_{\mu \nu }\left(\frac{\lambda (\phi)}{4}+ \lambda (\phi )
\partial_{\mu}\phi \partial^{\mu}\phi\right)=0\, .
\end{equation}
Finally, by using Eq. (\ref{lambdavar}), the equation (\ref{simlifiedeqnsmotion}) can be cast in the following form,
\begin{equation}\label{finaleqn}
\left (  \frac{g_{\mu \nu}R}{4}- R_{\mu
\nu}  \right )F'(R)-g_{\mu \nu }\left(\frac{3\lambda (\phi )}{4}\right)=0\, .
\end{equation}
By looking Eq. (\ref{finaleqn}), we conclude that since the first term is independent of $\phi$, the above equation can hold true in the following two cases:
\begin{itemize}
    \item Case I: Both the first and second term are equal to zero, that is,
    \begin{equation}\label{caseI}
\left (  \frac{g_{\mu \nu}R}{4}- R_{\mu
\nu}  \right )F'(R)=0,\,\,\,g_{\mu \nu }\left(\frac{3\lambda (\phi )}{4}\right)=0\, .
\end{equation}
    \item Case II: Both the first term are equal to the same constant, but with opposite signs, that is,
    \begin{equation}\label{crush}
\left (  \frac{g_{\mu \nu}R}{4}- R_{\mu
\nu}  \right )F'(R)=\Gamma,\,\,\,g_{\mu \nu }\left(\frac{3\lambda (\phi )}{4}\right)=-\Gamma\, ,
\end{equation}
where $\Gamma$ is a positive real matrix.
\end{itemize}    
In the following section, we shall analyze in detail the two cases we listed above, and we discuss in some detail the consequences corresponding to each of the two cases. We start off with case I, since it corresponds to the physical problem that interests us the most.

\subsubsection{Case I}

For this scenario let us analyze in some detail, what do the constraints (\ref{caseI}) imply for the vacuum mimetic $F(R)$ gravity model at hand. Firstly, the second constraint in Eq. (\ref{caseI}) can be true only if $\lambda (\phi)=0$, and by using Eq. (\ref{scalvar}), we easily conclude that the mimetic potential is, $V(\phi)=\Lambda$, with $\Lambda$ some arbitrary real constant, which we shall assume that is positive, without loss of generality. Hence, the allowed values of the mimetic potential $V(\phi )$ and of the Lagrange multiplier $\lambda (\phi)$, are given below,
\begin{equation}\label{defilen}
\lambda (\phi )=0,\,\,\,V(\phi )=\Lambda\, .
\end{equation}
Note that the solution (\ref{defilen}) results if we demand that the spherical symmetric metric with constant curvature of Eq. (\ref{metricressin}), is a solution of the vacuum mimetic $F(R)$ gravity with mimetic potential and Lagrange multiplier. The first constraint in Eq. (\ref{caseI}) is a bit more involved, so let us explicitly calculate the expression in order to have a clear picture of the implications that this constraint generates. By using the metric of Eq. (\ref{metricressin}), the first constraint explicitly reads,
\begin{align}\label{firstconstrainteqns}
\left(
\begin{array}{cccc}
 \frac{Q \left(-12 Q+r \left(12 M-12 r+r^3 R_0\right)\right)}{12 r^6} & 0 & 0 & 0 \\
 0 & -\frac{12 Q}{r^2 \left(-12 Q+r \left(12 M-12 r+r^3 R_0\right)\right)} & 0 & 0 \\
 0 & 0 & -\frac{Q}{r^2} & 0 \\
 0 & 0 & 0 & -\frac{Q \sin (\theta )^2}{r^2} \\
\end{array}
\right) F'(R_0)=0\, ,
\end{align}
which means that either $Q=0$ or $F'(R_0)=0$. Therefore, we have the following two scenarios,
\begin{itemize}
    \item Scenario I: This scenario corresponds to $Q\neq 0$, and therefore we have the following constraints corresponding to this scenario,
    \begin{equation}\label{scenarioI}
F'(R_0)=0,\,\,\,V(\phi )=\Lambda,\,\,\, \lambda (\phi)=0\, .
\end{equation}
    \item Scenario II: This scenario corresponds to $Q=0$ and it is described by the following constraints,
    \begin{equation}\label{scenarioI}
Q=0,\,\,\,V(\phi )=\Lambda,\,\,\, \lambda (\phi)=0\, .
\end{equation}
\end{itemize}    
Notice that in Scenario I, by using Eq. (\ref{finalfreqn}), we get, $F(R_0)=\Lambda$, while in Scenario II, we obtain that $F(R_0)=\frac{R_0F'(R_0)}{2}+\Lambda$. In Table \ref{TableI}, we gathered the results for the Scenarios I and II. In conclusion, the resulting picture of the mimetic $F(R)$ gravity, yields different results in comparison to the non-mimetic $F(R)$ gravity case studied in \cite{rnsergei1}. Particularly, in the present paper, the requirement that the AdS-RN black hole is a solution of the mimetic $F(R)$ gravitational system, results to many different cases for which this can be true, in comparison to the only cases $Q=0$ or $F'(R_0)=0$ corresponding to the case studied in the ordinary $F(R)$ gravity of Ref. \cite{rnsergei1}. This is easily explained, since the presence of the mimetic potential and of the Lagrange multiplier, offers more freedom in the resulting set of equations that need to be satisfied, namely Eqs. (\ref{simlifiedeqnsmotion}).
\begin{table*}[h]
\small
\caption{\label{TableI}The Scenarios I and II for the Mimetic $F(R)$ Gravity Reissner-Nordstr\"{o}m-anti de Sitter Black Hole}
\begin{tabular}{@{}crrrrrrrrrrr@{}}
\tableline
\tableline
\tableline
Scenario & Constraints $\,\,\,\,\,\,\,$$\,\,\,\,\,\,\,$$\,\,\,\,\,\,\,$$\,\,\,\,\,\,\,$$\,\,\,\,\,\,\,$$\,\,\,\,\,\,\,$$\,\,\,\,\,\,\,$
\\\tableline
Scenario I &  $F'(R_0)=0$,$\,\,\,\,\,\,\,$ $V(\phi )=\Lambda$, $\,\,\,\,\,\,\,$$\lambda (\phi)=0$, $\,\,\,\,\,\,\,$$F(R_0)=\Lambda$
\\\tableline
Scenario II & $\,\,\,\,\,\,\,$$\,\,\,\,\,\,\,$$\,\,\,\,\,\,\,$$F'(R_0)\neq 0$,$\,\,\,\,\,\,\,$ $V(\phi )=\Lambda$,$\,\,\,\,\,\,\,$ $\lambda (\phi)=0$,$\,\,\,\,\,\,\,$ $F(R_0)=\frac{R_0F'(R_0)}{2}+\Lambda$
\\
\tableline
\tableline
 \end{tabular}
\end{table*}
Before closing this section, we should discuss whether there exists another solution to Eq. (\ref{finaleqn}), different from the one described by Case I. It is worth discussing and studying this in detail, so by using the metric of Eq. (\ref{metricressin}) and inserting this in Eq. (\ref{finaleqn}), we obtain the following set of equations, which we quote in matrix form,
\begin{equation}\label{lastsamurai}
\left(
\begin{array}{cccc}
 \frac{\left(-12 Q+r \left(12 M-12 r+r^3 R_0\right)\right) \left(r^4 \lambda +Q F'(R_0)\right)}{12 r^6} & 0 & 0 & 0 \\
 0 & -\frac{12 \left(r^4 \lambda +Q F'(R_0)\right)}{r^2 \left(-12 Q+r \left(12 M-12 r+r^3 R_0\right)\right)} & 0 & 0 \\
 0 & 0 & r^2 \lambda -\frac{Q F'(R_0)}{r^2} & 0 \\
 0 & 0 & 0 & \frac{\sin (\theta )^2 \left(r^4 \lambda -Q F'(R_0)\right)}{r^2} \\
\end{array}
\right)=0\, .
\end{equation}
It is easy to see that the following two equations must simultaneously be satisfied, so that Eq. (\ref{lastsamurai}) holds true,
\begin{align}\label{lastsetofeqns}
&r^4 \lambda +Q F'(R_0)=0, \\ \notag &
r^4 \lambda -Q F'(R_0)=0\, .
\end{align}
The system of equations (\ref{lastsetofeqns}), has as solution what Case I describes, that is, $\lambda =0$ and $QF'(R_0)=0$, consequently this validates our claim that both terms of Eq. (\ref{finaleqn}), must independently be equal to zero.

 \subsubsection{Case II}

In this case, the constraints appearing in Eq. (\ref{crush}) must hold true. By observing Eq. (\ref{crush}), we can immediately see that neither $F'(R_0)$ and $Q$ can be zero, so for this case we have that,
\begin{equation}\label{caseiiconstraints}
F'(R_0)\neq 0,\,\,\,Q\neq 0\, .
\end{equation} 
Let us investigate whether the constraints of Eq. (\ref{crush}) can hold true. We start off with the second constraint, namely $g_{\mu \nu }\left(\frac{3\lambda (\phi )}{4}\right)=-\Gamma$, which by using the metric (\ref{metricressin}), it explicitly reads,
\begin{equation}\label{constraintforcaseii}
\left(
\begin{array}{cccc}
 \left(\frac{Q}{r^2}-\frac{M}{r}+\frac{1}{12} \left(12-r^2 R_0\right)\right) \lambda  & 0 & 0 & 0 \\
 0 & -\frac{\lambda }{\frac{Q}{r^2}-\frac{M}{r}+\frac{1}{12} \left(12-r^2 R_0\right)} & 0 & 0 \\
 0 & 0 & -r^2 \lambda  & 0 \\
 0 & 0 & 0 & -r^2 \lambda  \sin (\theta )^2 \\
\end{array}
\right)=\Gamma\, ,
\end{equation}
and we can easily see that the only constant solution for $\lambda$ is $\lambda=0$, and therefore $\Gamma=0$. Thereby, since $\Gamma=0$, the first constraint of Eq. (\ref{crush}) is satisfied when $Q=0$ or $F'(R_0)=0$, hence we end up to the first case, namely Case I.

As we demonstrated, only the Case I leads to a black hole solution for the vacuum mimetic $F(R)$ gravity with Lagrange multiplier and mimetic potential. The requirement that a AdS-RN black hole is a constant curvature solution of the mimetic $F(R)$ gravity, results to a certain number of constraints, which are different from the ordinary $F(R)$ gravity case studied in Ref. \cite{rnsergei1}. Therefore, it is natural to ask if these new conditions that the mimetic $F(R)$ gravity imposes, can have an effect on the perturbations of the AdS-RN black hole. In the sections to follow, we thoroughly address this question for both Scenario I, which corresponds to the AdS-RN black hole, since Scenario II, which occurs for $Q=0$, corresponds to the Schwarzschild-anti de Sitter black hole, and hence it is not of our interest for this paper, but we address this issue elsewhere.

\section{Perturbations of the Mimetic $F(R)$ AdS-RN Black Hole for Scenario I}

In this section, we perform the perturbations analysis for the Scenario I. Recall that in the context of Scenario I, the following constraints must be satisfied:
\begin{equation}\label{constrscenarioi}
F'(R_0)=0,\,\,\,V(\phi )=\Lambda,\,\,\, \lambda (\phi)=0,\,\,\, F(R_0)=\Lambda\, .
\end{equation}
As we already mentioned, the resulting picture is different from the ordinary $F(R)$ gravity studied in \cite{rnsergei1}, in which case, the function $F(R_0)$, satisfied $F(R_0)=0$, and of course, no potential term existed. The focus in this section is to investigate whether there is any difference in the perturbations of the AdS-RN, in the case of vacuum mimetic $F(R)$ gravity, when compared to the ordinary $F(R)$ gravity. In order to see this, we shall introduce some new coordinates and rewrite the metric of Eq. (\ref{metricressin}) in terms of these new coordinates. Particularly, we may redefine the radial variable $r$ in terms of the new variable $x$, and the parameter $r_1$ in terms of $r_0$, in the following way,
\begin{equation}\label{redefinitionofr}
r=r_0+\frac{\epsilon}{2}\left(1+\tanh x \right),\,\,\,r_1=r_0+\epsilon \, ,
\end{equation}
with $\epsilon$ a real positive parameter. Then, the function $A(r)$ appearing in Eq. (\ref{complementary}), can be expressed in terms of the variable $x$, as follows,
\begin{equation}\label{aform}
A(x)=-\frac{\epsilon^2}{4r_0^2}\left( 1-\frac{r_0^2R_0}{2}\right)\cosh^2x\, .
\end{equation}
In addition, we introduce the new time coordinate $\bar{t}$, which is related to the cosmic time coordinate $t$ in the following way, $\bar{t}=\frac{\epsilon \left( 1-\frac{r_0^2R_0}{2}\right)t}{2r_0^2}$\, ,
and by using this time variable redefinition and also Eq. (\ref{aform}), the AdS-RN metric of Eq. (\ref{metricressin}) receives the following form, 
\begin{equation}\label{metricredefinedpert}
\mathrm{d}s^2=\frac{r_0^2}{\left(1-\frac{r_0^2R_0}{2} \right)\cosh^2 x}\left(\mathrm{d}\bar{t}^2-\mathrm{d}x^2 \right)+r_0^2\mathrm{d}\Omega^2\, .
\end{equation}
By introducing the variables $M_+$ and $M_-$, which are defined as follows,
\begin{equation}\label{metricnewvariablesbandc}
M_{+}=\frac{\sqrt{1-\frac{r_0^2R_0}{2}}}{r_0},\,\,\, M_{-}=\frac{1}{r_0}\, ,
\end{equation}
the metric of Eq. (\ref{metricredefinedpert}) receives the following simplified form,
\begin{equation}\label{metricperturbed}
\mathrm{d}s^2=\frac{1}{M_{+}^2\cosh^2 x}\left(\mathrm{d}\bar{t}^2-\mathrm{d}x^2 \right)+\frac{1}{M_{-}^2}\mathrm{d}\Omega^2\, .
\end{equation}
We perform the following perturbation on the metric of Eq. (\ref{metricperturbed}),
\begin{equation}\label{metricperturbed1}
\mathrm{d}s^2=\frac{e^{2\rho(x,\bar{t})}}{M_{+}^2\cosh^2 x}\left(\mathrm{d}\bar{t}^2-\mathrm{d}x^2 \right)+\frac{e^{-2\varphi(x,\bar{t})}}{M_{-}^2}\mathrm{d}\Omega^2\, ,
\end{equation}
with the functions $\rho (x,\bar{t})$ and $\varphi (x,\bar{t})$ being smooth and differentiable functions of $x$ and $\bar{t}$, and being chosen in the following way,
\begin{equation}\label{functionsperturbations}
\rho (x,\bar{t})=-\ln (\cosh x)+\delta \rho,\,\,\, \varphi=\delta \varphi\, .
\end{equation} 
We can easily compute the effect of the perturbed metric (\ref{metricperturbed1}), if we use the mimetic $F(R)$ gravity equations of motion of Eq. (\ref{equationofmotionbasicconstantcurv}), which in view of the constraints (\ref{constrscenarioi}), receive the following form,
\begin{align}\label{equationsmotionfinalscenarioi}
& \frac{1}{2}g_{\mu \nu}F(R)-R_{\mu
\nu}F'(R)-
\frac{1}{2}g_{\mu \nu}\Lambda =0 \, .
\end{align}
In order to study the effect of the perturbed metric appearing in Eq. (\ref{metricperturbed1}) on the equations of motion (\ref{equationsmotionfinalscenarioi}), we expand Eq. (\ref{equationsmotionfinalscenarioi}) in terms of the components of the metric, and the result is,
\begin{align}\label{comp1}
&\frac{e^{2\rho}}{2M_+^2}F(R)-\Big{(}\ddot{\rho}+2\ddot{\varphi}+\frac{\partial^2\rho}{\partial x^2}-2\dot{\varphi}^2-\frac{\partial\rho}{\partial x}\frac{\partial \varphi}{\partial x }-2\dot{\rho}\dot{\varphi}\Big{)}F'(R)+\frac{\partial^2F'(R)}{\partial \bar{t}^2}-\dot{\rho}\frac{\partial F'(R)}{\partial \bar{t}}-\frac{\partial \rho}{\partial x}\frac{\partial F'(R)}{\partial x}\\ \notag &+e^{2\varphi }\frac{\partial }{\partial x }\Big{(}e^{-2\varphi }\frac{\partial F'(R)}{\partial x}\Big{)}-e^{2\varphi }\frac{\partial }{\partial \bar{t} }\Big{(}e^{-2\varphi }\frac{\partial F'(R)}{\partial \bar{t}}\Big{)}-\frac{e^{2\rho}}{2M_+^2}\Lambda=0\, ,
\end{align}
\begin{align}\label{comp2}
&-\frac{e^{2\rho}}{2M_+^2}F(R)-\Big{(}\ddot{\rho}+2\frac{\partial^2\varphi }{\partial x^2}-\frac{\partial^2\rho}{\partial x^2}-2\left(\frac{\partial\rho}{\partial x}\right)^2-2\frac{\partial\rho}{\partial x}\frac{\partial\varphi}{\partial x}-2\dot{\rho}\dot{\varphi}\Big{)}F'(R)+\frac{\partial^2F'(R)}{\partial x^2}-\dot{\rho}\frac{\partial F'(R)}{\partial \bar{t}}-\frac{\partial \rho}{\partial x}\frac{\partial F'(R)}{\partial x}\\ \notag &-e^{2\varphi }\frac{\partial }{\partial x }\Big{(}e^{-2\varphi }\frac{\partial F'(R)}{\partial x}\Big{)}-e^{2\varphi }\frac{\partial }{\partial \bar{t} }\Big{(}e^{-2\varphi }\frac{\partial F'(R)}{\partial \bar{t}}\Big{)}+\frac{e^{2\rho}}{2M_+^2}\Lambda=0\, ,
\end{align}
\begin{align}\label{comp3}
& -\left( -2\frac{\partial \dot{\varphi} }{\partial x }-2 \frac{\partial \varphi }{\partial x }\dot{\varphi}-2 \frac{\partial \rho }{\partial x }\dot{\varphi}-2\dot{\rho}\frac{\partial \varphi}{\partial x }\right)F'(R)+\frac{\partial^2F'(R) }{\partial x \partial \bar{t} }-\dot{\rho}\frac{\partial F'(R)}{\partial x } -\frac{\partial \rho }{\partial x }\frac{\partial F'(R)}{\partial \bar{t} }=0\, ,
\end{align}
\begin{align}\label{comp4}
& \frac{e^{-2\varphi}}{2M_-^2}F(R)-\frac{M_+^2}{M_-^2}e^{-2(\rho+\varphi)}\left ( -\ddot{\varphi}+2 \frac{\partial^2 \varphi }{\partial x^2}-2\left(\frac{\partial \varphi}{\partial x}\right)^2+2\dot{\varphi}^2\right) F'(R)+F'(R)
\\ \notag & +\frac{M_+^2}{M_-^2}e^{-2(\rho+\varphi)}\left ( \dot{\varphi}\frac{\partial F'(R) }{\partial \bar{t}}-\frac{\partial \varphi}{\partial x}\frac{\partial F'(R)}{\partial x}\right)-\frac{M_+^2}{M_-^2}e^{-2\rho}\Big{[}-\frac{\partial }{\partial \bar{t}}\left(e^{-2\varphi }\frac{\partial F'(R)}{\partial \bar{t} }\right)+ \frac{\partial }{\partial x}\left(e^{-2\varphi }\frac{\partial F'(R)}{\partial x }\right)\Big{]}-\frac{e^{-2\varphi}}{2M_-^2}\Lambda\, ,
\end{align}
where Eq. (\ref{comp1}) corresponds to the $(\bar{t},\bar{t})$ component, Eq. (\ref{comp2}) to the $(x,x)$ component,  Eq. (\ref{comp3}) to the $(x,\bar{t})$ and $(\bar{t},x)$ components and Eq. (\ref{comp4}) to the $(\theta,\theta)$ and $(\phi ,\phi )$ components. Notice that Eq. (\ref{comp3}) does not have any contribution from the mimetic potential, and also note that due to the form of the Riemann tensor and of the Christoffel symbols of the metric (\ref{metricperturbed1}), the equations of motion corresponding to the $(\theta,\theta)$ and $(\phi ,\phi )$ components are identical. We can see how the perturbations affect the equations of motion, by considering the variation of each equation above, around $R_0$, and by keeping first order terms. We shall present in detail the variation of the first term, and the rest can be done accordingly. By varying the first term of Eq. (\ref{comp1}), we have,
\begin{equation}\label{var1}
\delta \left[ \frac{e^{2\rho}}{2M_+^2}F(R)\right]=\frac{\delta \rho e^{2\rho}}{2M_+^2}F(R_0)+\frac{e^{2\rho}}{2M_+^2}F'(R_0)\delta R\, ,
\end{equation}
and since from the constraints of Scenario I (\ref{constrscenarioi}), we have that $F'(R_0)=0$, we finally get that,
\begin{equation}\label{dsfer}
\delta \left[ \frac{e^{2\rho}}{2M_+^2}F(R)\right]=\frac{\delta \rho e^{2\rho}}{2M_+^2}F(R_0)\, .
\end{equation}
Accordingly, by noticing that $\ddot{\rho}\sim \delta \rho$ and $ \ddot{\varphi}\sim \delta \varphi $, and in addition that,
\begin{equation}\label{eauxil}
\frac{\partial^2 \rho}{\partial x^2}=\frac{1}{\cosh ^2 x}+\frac{\partial^2 \delta \rho}{\partial x^2}\, ,
\end{equation}
and by keeping only first order terms, the variation of the second term of Eq. (\ref{comp1}) yields,
\begin{equation}\label{dfedd}
\delta \Big{(}-2\dot{\varphi}^2-\frac{\partial\rho}{\partial x}\frac{\partial \varphi}{\partial x }-2\dot{\rho}\dot{\varphi}\Big{)}F'(R))\simeq -\frac{1}{\cosh^2 x}F''(R_0)\delta R \, .
\end{equation}
Finally, upon varying the third, fourth, fifth and sixth term of Eq. (\ref{comp1}), we obtain the following results,
\begin{align}\label{firstorederannalyticvariations}
& \delta \left[ \frac{\partial^2F'(R)}{\partial \bar{t}^2}\right ]=F''(R_0)\ddot{\delta R}\, ,\\ \notag &
\delta \left[ -\dot{\rho}\frac{\partial F'(R)}{\partial \bar{t}}\right ]=-\dot{\rho}F''(R_0)\delta \dot{\rho}\, ,\\ \notag &
\delta \left[ -\frac{\partial \rho}{\partial x}\frac{\partial F'(R)}{\partial x}\right ]=\tanh x F''(R_0) \frac{\partial\delta R}{\partial x }\, ,\\ \notag &
\delta \left[ e^{2\varphi }\frac{\partial }{\partial x }\Big{(}e^{-2\varphi }\frac{\partial F'(R)}{\partial x}\Big{)}\right ]=-F''(R_0)\ddot{\delta R}\, , \\ \notag &
\delta \left[ -e^{2\varphi }\frac{\partial }{\partial \bar{t} }\Big{(}e^{-2\varphi }\frac{\partial F'(R)}{\partial \bar{t}}\Big{)}\right ]=F''(R_0) \frac{\partial^2 \delta R}{\partial x^2 }\, , \\ \notag &
\delta \left[ -\frac{e^{2\rho}}{2M_+^2}\Lambda\right]=-\frac{e^{2\rho}}{2M_+^2}\Lambda\delta \rho\, ,
\end{align}
where we kept only first order terms. Notice that the fourth term is of second order, so it is omitted eventually. At this point we can identify the differences between the mimetic $F(R)$ gravity and the ordinary $F(R)$ gravity studied in \cite{rnsergei1}. By a direct comparison, in the ordinary $F(R)$ case, the variation of the first term appearing in Eq. (\ref{dsfer}) yields that,
\begin{equation}\label{dsferanew}
\delta \left[ \frac{e^{2\rho}}{2M_+^2}F(R)\right]=\frac{\delta \rho e^{2\rho}}{2M_+^2}F(R_0)=0\, ,
\end{equation}
since $F(R_0)=0$ for the ordinary $F(R)$ gravity. On the contrary, in the mimetic $F(R)$ case, the term $F(R_0)$ is not zero, but equal to $F(R_0)=\Lambda $, and hence the difference between the two cases is obvious. However, by replacing $F(R_0)=\Lambda $ in Eq. (\ref{dsfer}), we can see that it is cancelled by the last term of Eq. (\ref{firstorederannalyticvariations}), which originates from the mimetic potential. By combining Eqs. (\ref{dsfer}) and (\ref{firstorederannalyticvariations}), we obtain the following perturbed equation,
\begin{equation}\label{perturbedeq1}
F''(R_0)\Big{[} -\frac{1}{\cosh^x}\delta R+\tanh x\frac{\partial \delta R}{\partial x } +\frac{\partial^2 \delta R}{\partial x^2 }\Big{]}=0\, ,
\end{equation}
Accordingly, by perturbing Eqs. (\ref{comp2}), (\ref{comp3}) and (\ref{comp4}), we obtain the following perturbation equations,
\begin{equation}\label{perturbedeq2}
F''(R_0)\Big{[} \frac{1}{\cosh^x}\delta R+\delta\ddot{ R} +\tanh x\frac{\partial \delta R}{\partial x }\Big{]}=0\, ,
\end{equation}
\begin{equation}\label{perturbedeq3}
F''(R_0)\Big{[} \frac{\partial \delta \dot{R}}{\partial x } +\tanh x\delta \dot{R}\Big{]}=0\, ,
\end{equation}
\begin{equation}\label{perturbedeq4}
F''(R_0)\Big{[} \delta R+\cosh^2 x\left(-\delta \ddot{R} +\frac{\partial^2 \delta R}{\partial x^2 }\right)\Big{]}=0\, ,
\end{equation}
A direct comparison of the perturbation equations (\ref{perturbedeq1}), (\ref{perturbedeq2}), (\ref{perturbedeq3}) and (\ref{perturbedeq4}), to the ones corresponding to non-mimetic $F(R)$ gravity in Ref. \cite{rnsergei1}, shows that the two cases lead to the same perturbation equations. Therefore, although in the case of mimetic $F(R)$ gravity, the conditions under which the AdS-RN black holes is a solution of the gravitational equations, are different from the ordinary $F(R)$ gravity case, the perturbation equations around the constant curvature solution are identical to the ones corresponding to ordinary $F(R)$ gravity. Correspondingly, one expects that the anti-evaporation phenomena that occurred for the ordinary $F(R)$ AdS-RN black holes, will also hold true in the mimetic $F(R)$ case too. So let us investigate this possibility in some detail. We start off by calculating the variation of the Ricci scalar corresponding to the metric of Eq. (\ref{metricperturbed1}), which we denote as $\delta R$. In order to see how $\delta R$ looks like at first order with regards to $\delta \rho $ and $\delta \varphi$, we need to explicitly calculate the Ricci scalar corresponding to the metric (\ref{metricperturbed1}), which is,
\begin{align}\label{explicitlyricciscalar}
& R=2 e^{-2 \rho [x,t]}\Big{(}e^{2 (\rho [x,t]+\varphi [x,t])}M_-^2-3 M_+^2\left(\frac{\partial \varphi }{\partial \bar{t}}\right)^2-M_+^2\frac{\partial^2 \rho }{\partial \bar{t}^2}+2M_+^2\frac{\partial^2 \varphi }{\partial \bar{t}^2}+3M_+^2\left(\frac{\partial \varphi }{\partial x}\right)^2+M_+^2\frac{\partial^2 \rho }{\partial x^2}-2M_+^2\frac{\partial^2 \varphi }{\partial x^2}
 \Big{)}\, .
\end{align} 
Varying the above equation, the variation $\delta R$ at first order approximation of $\delta \rho$ and $\delta \varphi$ and their corresponding derivatives, is equal to,
\begin{equation}\label{olonsygkrotei}
\delta R=- M_+^2\delta \rho+4M_-^2\delta \varphi-M_+^2\cosh ^2 x\left( 2\left(\delta\ddot{\rho }-\frac{\partial^2\delta \rho}{\partial x^2} \right)-4\left(\delta \ddot{\varphi}-\frac{\partial^2\delta \varphi}{\partial x^2} \right)\right)\, .
\end{equation}
From the perturbation equations (\ref{perturbedeq1})-(\ref{perturbedeq4}), by assuming that $F''(R_0)\neq 0$ and by combining (\ref{perturbedeq1}) and (\ref{perturbedeq2}), we obtain,
\begin{equation}\label{anotherrelaimissthes}
\delta R=-\frac{\cosh^2x}{2}\left( \frac{\partial^2\delta R}{\partial x^2}-\delta \ddot{R}\right)\, ,
\end{equation}
while Eq. (\ref{perturbedeq4}), dictates that,
\begin{equation}\label{anotherrelaimissthes1}
\delta R=-\cosh^2x\left( \frac{\partial^2\delta R}{\partial x^2}-\delta \ddot{R}\right)\, .
\end{equation}
Therefore, the combination of Eqs. (\ref{anotherrelaimissthes}) and (\ref{anotherrelaimissthes1}), yields that $\delta R=0$. By assuming the parametrization $\delta \rho=\rho_0\cosh^{\beta} (\omega \bar{t})$, $\delta \varphi =\varphi_0\cosh^{\beta} (\omega \bar{t})$ \cite{rnsergei1}, where $\omega$, $\beta$, $\varphi_0$ and $\rho_0 $ being arbitrary parameters, and combining Eq. (\ref{olonsygkrotei}) with the parametrization for $\delta \rho$ and $\delta \varphi$, we obtain the following equation,
\begin{equation}\label{asxetieqn}
\left(M_+^2\rho_0-2M_-^2\varphi_0\right)(\omega^2-\beta^2)=0\, .
\end{equation}
In order to obtain a non-trivial solution for the perturbation equations, we conclude that Eq. (\ref{asxetieqn}) holds true only when $\omega=\beta $, in which case, $\varphi_0$ and one of $\omega$ or $\beta$ can be arbitrary. Hence if $\varphi_0<0$, the perturbation of the metric is unstable, since $\delta \varphi =\varphi$, grows with time. Before closing, we must mention that this instability was interpreted in Ref. \cite{rnsergei1}, as anti-evaporation of the AdS-RN black hole.

In conclusion, the vacuum mimetic $F(R)$ gravity has the AdS-RN black hole as a solution, but different conditions must hold true in order for this to be true, compared to the ordinary $F(R)$ gravity case. However, the perturbations of the AdS-RN black hole in the mimetic $F(R)$ case, are not affected by the presence of the mimetic Lagrange multiplier or the mimetic scalar potential. Therefore, the instabilities, which could be interpreted as anti-evaporation phenomena, persist in the mimetic $F(R)$ case too.

Before we close this section, let us briefly speculate why the perturbation equations in the mimetic $F(R)$ gravity, for the AdS-RN black hole, are the same in comparison to the ordinary $F(R)$ gravity case. Firstly, a reason for this could be that the perturbation equations are actually first order in the perturbed variables. Therefore, this coincidence may be an artifact of the first order linear approximation for the black hole perturbation. A second, most possible reason behind this feature, is that the coincidence of the perturbation equation is due to the specific form of the AdS-RN black hole. Actually, the condition $F'(R_0)=0$, simplifies a lot the equations in the mimetic case too, hence if $F'(R_0)\neq 0$, it is possible that the perturbation equations in the mimetic and ordinary $F(R)$ gravity cases, are in fact different. This may be the case in the Schwarzschild de Sitter black hole, which was studied in the context of the $F(R)$ gravity in Ref. \cite{rnsergei2}. We hope to address this issue in a future work.

\section{Conclusions}

In this paper we studied the AdS-RN black hole, in the context of mimetic $F(R)$ gravity with Lagrange multiplier and mimetic potential. As we demonstrated, by imposing the condition that the AdS-RN black hole is a solution of the mimetic $F(R)$ gravity, results to some constraints, which are different in comparison to the ones corresponding to the ordinary $F(R)$ gravity case. Particularly, in Table \ref{doxa} below, we gather the differences in the constraints that need to be satisfied, but also the similarities, of the ordinary and mimetic vacuum $F(R)$ gravity cases. 
\begin{table*}[h]
\small
\caption{\label{doxa}Comparison of the Constraints Corresponding to the vacuum Mimetic $F(R)$ Gravity and to the vacuum Ordinary $F(R)$ Gravity for the Reissner-Nordstr\"{o}m-anti de Sitter Black Hole}
\begin{tabular}{@{}crrrrrrrrrrr@{}}
\tableline
\tableline
\tableline
$F(R)$ Model & Constraints $\,\,\,\,\,\,\,$$\,\,\,\,\,\,\,$$\,\,\,\,\,\,\,$ &$\,\,\,\,\,\,\,$ $\,\,\,\,\,\,\,$$\,\,\,\,\,\,\,$$\,\,\,\,\,\,\,$$\,\,\,\,\,\,\,$$\,\,\,\,\,\,\,$$\,\,\,\,\,\,\,$$\,\,\,\,\,\,\,$$\,\,\,\,\,\,\,$Anti-Evaporation
\\\tableline
Mimetic $F(R)$ Gravity & $\,\,\,\,\,\,\,$$\,\,\,\,\,\,\,$$\,\,\,\,\,\,\,$$F(R_0)=\Lambda$, $F'(R_0)=0$, $\lambda=0$, $V(\phi)=\Lambda$  & $\,\,\,\,\,\,\,$ Yes
\\\tableline
Ordinary $F(R)$ Gravity & $\,\,\,\,\,\,\,$$\,\,\,\,\,\,\,$$\,\,\,\,\,\,\,$$F(R_0)=0$, $F'(R_0)=0$, $\lambda=0$, $V(\phi)=0$  & $\,\,\,\,\,\,\,$ Yes
\\\tableline
\tableline
 \end{tabular}
\end{table*}
In addition, we also investigated how the perturbations of the AdS-RN black hole are affected by the presence of the Lagrange multiplier and mimetic potential. As we explicitly demonstrated, the result is quite interesting, since the perturbation equations at first order are the same, however, differences may appear at higher orders in the perturbation variables. As we discussed earlier, it seems that the structure of the first order perturbations are the same due to the form of the AdS-RN black hole. Particularly, the requirement that the AdS-RN black hole is a solution of the mimetic $F(R)$ gravity, results in the constraint that $F'(R_0)=0$, and this constraint is the reason why the perturbations equations are the same in the cases under study. Finally, as in the ordinary $F(R)$ case, anti-evaporation occurs in the mimetic $F(R)$ AdS-RN black hole too, since the perturbations become unstable in this case too.

Motivated by the fact that, the reason for having the same perturbations equations for the mimetic and ordinary $F(R)$ AdS-RN black hole, is the constraint $F'(R_0)=0$, it is interesting and tempting to study black holes for which the constraint $F'(R_0)=0$ no longer applies. This for example could be the case for the charged black hole studied in \cite{rnsergei1}, or in the case of the Schwarzschild-de Sitter black hole, studied in \cite{rnsergei2}. It would therefore be quite interesting to study these cases too, which we hope to address in a future publication.

Another interesting possibility for further study is related to primordial black holes, since the black holes we studied in this paper are expected to be primordial black holes. Recently, an interesting study related to non-Gaussianities and the formation of supermassive black holes was performed in Ref. \cite{brande}. It is interesting to see if these non-Gaussianities could have some possible effect on the evolution of primordial black holes. This for example could be combined with the study of gravitational memory of the primordial black holes \cite{barrow}, an effect which could have an imprint on the primordial black holes at the time of their formation. For a recent study on gravitational memory in the context of $F(R)$ gravity, see \cite{gravioik}.

Finally, since the instability of the AdS-RN black hole, provides the holographic description of the phase transitions that occur in the dual condensed matter theory \cite{holo2,holo3} of an Einstein-Maxwell gravitational theory, it would be interesting to see if there is any possible connection between the $F(R)$ AdS-RN black hole, and the condensed matter systems, like superfluid or superconducting systems. Notice that the most interesting feature is that in the context of $F(R)$ gravity, no abelian Maxwell fields are needed in order for the AdS-RN black hole to be a solution. We hope to address some of these issues in the future.

\section*{Acknowledgments}

This work is supported by Min. of Education and Science of Russia (V.K.O).

\section*{Appendix: The Parameters $M$ and $Q$ in Terms of $r_0$ and $r_1$}

Here we quote the exact form of the parameters $M$ and $Q$ appearing in the AdS-RN black hole metric of Eq. (\ref{metricressin}). Particularly, the parameter $Q$ is assumed to have the following form,
\begin{equation}\label{parameterQ}
Q=r_0r_1\left( 1-\frac{\left(r_0^2+r_1^2+r_0r_1\right)R_0}{12}\right)\, ,
\end{equation}
and the parameter $M$ is assumed to be,
\begin{equation}\label{parameterMappendix}
M=\left( r_0+r_1\right)\left( 1-\frac{\left( r_0^2+r_1^2\right)R_0}{12}\right)\, .
\end{equation}
By choosing the mass parameter $M$ and the ``charge'' parameter $Q$ as in Eqs. (\ref{parameterQ}) and (\ref{parameterMappendix}), the two horizons of the AdS-RN black hole occur at $r_0$ and $r_1$, and the function $A(r)$ can take the form appearing in Eq. (\ref{complementary}).

\end{document}